\documentclass[twocolumn,showpacs,amsmath,amssymb,aps,
prd,nofootinbib]{revtex4}

\usepackage{graphicx}

\begin{document}

\title{The influence of differential rotation on the detectability of
gravitational waves \\ from the \emph{r}-mode instability}

\author{Paulo M. S\'a}

\email{pmsa@ualg.pt}

\author{Brigitte Tom\'e}

\email{btome@ualg.pt}

\affiliation{Centro Multidisciplinar de Astrof\'{\i}sica --
CENTRA, Departamento de F\'{\i}sica,  Faculdade de Ci\^encias e
Tecnologia, Universidade do Algarve, Campus de Gambelas, 8005-139
Faro, Portugal}

\date{May 31, 2006}

\begin{abstract}
Recently, it was shown that differential rotation is an
unavoidable feature of nonlinear \emph{r}-modes. We investigate
the influence of this differential rotation on the detectability
of gravitational waves emitted by a newly born, hot,
rapidly-rotating neutron star, as it spins down due to the
\emph{r}-mode instability. We conclude that gravitational
radiation may be detected by the advanced laser interferometer
detector LIGO if the amount of differential rotation at the time
the \emph{r}-mode instability becomes active is not very high.
\end{abstract}


\maketitle

\section{\label{sect-introduction}Introduction}

The new kilometer-scale laser interferometer gravitational-wave
detectors LIGO and Virgo, or their advanced versions, operating at
a broad frequency band (between about 10 and $10^4$ Hz), may
detect gravitational waves from a variety of sources (for a recent
review see Ref.~\cite{KS2006}). One such source could be a
spinning neutron star emitting gravitational radiation due to the
\emph{r}-mode instability.

First studied more than twenty years ago \cite{PP1978},
\emph{r}-modes are non-radial oscillation modes of rotating stars
that have the Coriolis force as their restoring force and a
characteristic frequency comparable to the angular velocity of the
star. These modes have been attracting increasing attention since
it was discovered that they are driven unstable by gravitational
radiation emission \cite{A1998,FM1998} and that, for a large range
of relevant temperatures and angular velocities of neutron stars,
the driving effect of gravitational radiation is stronger than the
damping effect of viscosity \cite{LOM1998,AKS1999}.

Using a phenomenological model for the evolution of the
\emph{r}-mode instability, Owen \emph{et al.}\ \cite{OLCSVA1998}
studied the detectability of gravitational waves emitted by newly
born, hot, rapidly-rotating neutron stars, arriving at the
conclusion that such waves could be detected by the enhanced
version of LIGO if sources were located at distances up to 20 Mpc
from Earth. However, a deeper understanding of this issue requires
taking into account nonlinear effects in the evolution of the
\emph{r}-mode instability. Arras \emph{et al.}\ \cite{AFMSTW2003}
considered the nonlinear interaction between modes, arriving at
the conclusion that enhanced laser interferometer detectors could
see gravitational radiation from \emph{r}-modes if the sources
were much closer to Earth, namely, at distances smaller than about
200 kpc. In this paper we will investigate the issue of
detectability of gravitational radiation emitted by newly born
neutron stars as they spin down due to the \emph{r}-mode
instability, but taking into account another nonlinear effect:
differential rotation.

Rezzolla \emph{et al.}\ \cite{RLS2000} were the first to suggest
that \emph{r}-modes induce a drift of fluid elements along
azimuthal directions and derived an approximate analytical
expression for these drifts. Soon afterwards, numerical studies,
both in general relativistic \cite{SF2001} and Newtonian
hydrodynamics \cite{LTV2001}, confirmed the existence of such
drifts. Recently, an exact \emph{r}-mode solution, representing
differential rotation of kinematic nature that produce large scale
drifts of fluid elements along stellar latitudes, was found within
the nonlinear theory up to second order in the mode's amplitude in
the case of a Newtonian, barotropic, nonmagnetized, perfect-fluid
star \cite{Sa2004}. This differential rotation, which is an
unavoidable feature of nonlinear \emph{r}-modes, contributes to
the physical angular momentum of the mode \cite{Sa2004} and,
therefore, plays an important role in the nonlinear evolution of
the \emph{r}-mode instability \cite{ST2005}. In particular, the
amplitude of the \emph{r}-mode saturates in a natural way at a
value that depends on the amount of differential rotation at the
time the instability becomes active; this saturation value can be
much smaller than unity \cite{ST2005}.

In this paper, using the model of evolution of Ref.~\cite{ST2005},
we investigate how differential rotation induced by \emph{r}-modes
influences the detectability, by the laser interferometer
detectors LIGO and Virgo, of gravitational waves emitted by a
newly born, hot, rapidly-rotating neutron star. In
Sect.~\ref{sect-role-dif-rot} we briefly review the evolution
model of the \emph{r}-mode instability. The main results of the
paper are presented in Sect.~\ref{sect-detectability}, where we
derive expressions for the maximum value of the gravitational wave
amplitude and for the frequency-domain gravitational waveform. We
also compare, for different values of differential rotation, the
characteristic amplitude of the signal with the rms strain noise
in the initial LIGO, Virgo and advanced LIGO gravitational-wave
detectors and compute the signal-to-noise ratio. Finally,
Sect.~\ref{sect-conclusions} is devoted to a discussion of the
results and conclusions.

\section{\label{sect-role-dif-rot} Nonlinear evolution of the
\emph{r}-mode instability}

In a recent paper \cite{ST2005}, we have investigated the role of
differential rotation in the evolution of the $l=2$ \emph{r}-mode
instability in a newly born, hot, rapidly-rotating neutron star,
using a simple phenomenological model adapted from the one
proposed in Ref.~\cite{OLCSVA1998}. The main difference between
our modified model and the one proposed in Ref.~\cite{OLCSVA1998}
is that the former takes into account that the full physical
angular momentum of the \emph{r}-mode perturbation also includes a
contribution from differential rotation. Namely, at second order
in the mode's amplitude $\alpha$, the physical angular momentum
was taken to be \cite{Sa2004}
\begin{eqnarray}
\delta^{(2)}\! J = \frac12 \alpha^2 \Omega (4K+5) \tilde{J} M R^2,
 \label{phys-ang-mom}
\end{eqnarray}
where $\Omega$, $R$ and $M$ denote the angular velocity, the
radius and the mass of the star, respectively, and
$\tilde{J}\equiv\int_0^R \rho r^6 dr/(MR^4)=1.635\times 10^{-2}$.
In the above expression, it was assumed that the star's mass
density $\rho$ and pressure $p$ are related by a polytropic
equation of state $p=k\rho^2$ with $k$ such that $M=1.4 M_{\odot}$
and $R=12.53 \mbox{ km}$. In Eq.~(\ref{phys-ang-mom}), $K$ is a
constant fixed by initial data, giving the initial amount of
differential rotation associated with the \textit{r}-mode.

Within our model (see Ref.~\cite{ST2005} for details), the total
angular momentum is given by the sum of the angular momentum of
the unperturbed star and the angular momentum of the \emph{r}-mode
perturbation,
\begin{eqnarray}
J=I\Omega+\delta^{(2)} \! J(\Omega,\alpha),
 \label{total-ang-mom}
\end{eqnarray}
where $I=(8\pi/3)\int_0^R\rho r^4 dr=\tilde{I}MR^2$
($\tilde{I}=0.261$) is the momentum of inertia of the unperturbed
star. Assuming that the total angular momentum of the star
decreases due to the emission of gravitational radiation and that
the angular momentum of the perturbation increases due to the
emission of gravitational radiation and decreases due to the
dissipative effect of viscosity, we arrive at a system of
differential equations determining the evolution of the star's
angular velocity $\Omega(t)$ and the \emph{r}-mode's amplitude
$\alpha(t)$, namely,
\begin{eqnarray}
\frac{d\Omega}{dt} &=& \frac83 (K+2)Q \alpha^2
\frac{\Omega}{\tau_{GR}} + \frac83 \left( K+\frac54 \right)Q
\alpha^2 \frac{\Omega}{\tau_{V}}, \label{eq-dif-omega(t)-c-visc}
\\
\frac{d\alpha}{dt} &=& -\left[ 1 + \frac43 (K+2)Q \alpha^2 \right]
\frac{\alpha}{\tau_{GR}} \nonumber \\
& & - \left[ 1 + \frac43 \left( K+\frac54 \right)Q \alpha^2
\right] \frac{\alpha}{\tau_{V}}, \label{eq-dif-alpha(t)-c-visc}
\end{eqnarray}
where $Q\equiv3\tilde{J}/(2\tilde{I})=0.094$ and the value of $K$
is chosen to lie in the interval $-5/4\leqslant K\ll 10^{13}$. The
upper limit for $K$ results from the fact that one wishes to
impose the condition that the initial value of the angular
momentum of the \emph{r}-mode is much smaller than the angular
momentum of the unperturbed star, i.e., $\delta^{(2)}\!J_0 \ll
I\Omega_0$ implies that $K\ll 10^{13}$ if we choose
$\alpha_0=10^{-6}$. The lower limit for $K$ results from the fact
that we do not want to saturate the amplitude of the mode by hand,
a procedure needed for the case $K<-5/4$ in order to avoid that
the total angular momentum of the star becomes negative (see
Ref.~\cite{ST2005} for details).

In Eqs.~(\ref{eq-dif-omega(t)-c-visc}) and
(\ref{eq-dif-alpha(t)-c-visc}) the gravitational-radiation and
viscous timescales\footnote{The gravitational-radiation timescale
$\tau_{GR}$ for \emph{r}-modes was first obtained using an
expression for the time evolution of the physical energy of the
\emph{r}-mode perturbation, assuming that the imaginary part of
the frequency of the mode, $Im(\omega)\equiv 1/\tau_{GR}$, is
related to the time derivative of the energy by
$dE/dt=-2E/\tau_{GR}$ \cite{LOM1998}. But $\tau_{GR}$ can also be
obtained by solving \emph{explicitly} the hydrodynamic equations
in the presence of the gravitational radiation reaction force
\cite{DS2005}.} are given, respectively, by \cite{LOM1998}
\begin{eqnarray}
\frac{1}{\tau_{GR}} &=& \frac{1}{\tilde{\tau}_{GR}} \left(
\frac{\Omega}{ \sqrt{\pi G \bar{\rho}} } \right)^6,
\label{gr-timescale}
\\
\frac{1}{\tau_V} &=& \frac{1}{\tilde{\tau}_S} \left(
\frac{10^9\mbox{K}}{T} \right)^2 + \frac{1}{\tilde{\tau}_B} \left(
\frac{T}{10^9\mbox{K}} \right)^6 \left( \frac{\Omega}{\sqrt{\pi G
\bar{\rho}}} \right)^2 \! \!, \label{visc-timescale}
\end{eqnarray}
with the fiducial timescales $\tilde{\tau}_{GR}=-3.26 \mbox{ s}$,
$\tilde{\tau}_S = 2.52\times10^8 \mbox{ s}$ \cite{LOM1998} and
$\tilde{\tau}_B=2.01\times10^{11} \mbox{ s}$ \cite{LMO1999}. The
temperature of the star is assumed to decrease due to the emission
of neutrinos via a modified URCA process,
\begin{eqnarray}
\frac{T(t)}{10^9\mbox{ K}}=\left[ \frac{t}{\tau_c} +\left(
\frac{10^9\mbox{ K}}{T_0} \right)^6 \right]^{-1/6}, \label{T(t)}
\end{eqnarray}
where $T_0$ is the initial temperature of the star and
$\tau_c=1\mbox{ yr}$ characterizes the cooling rate
\cite{OLCSVA1998}.

The viscous timescale given above was derived for a simple model
of a neutron star with shear and bulk viscosity. The consideration
of other types of viscosity (as, for instance, viscosity at the
core-crust boundary in neutron stars with a crust \cite{BU2000} or
hyperon bulk viscosity \cite{LO2002}) leads to expressions for the
viscous timescale that differ substantially from
Eq.~(\ref{visc-timescale}). However, because of the complexity of
the involved processes it is not clear yet which viscous timescale
is more appropriate to describe a real neutron star. In this paper
we choose to restrict ourselves to the viscous timescale given by
Eq.~(\ref{visc-timescale}).

Let us assume that the temperature, the mode's amplitude and the
star's angular velocity take the initial values $T_0=10^{11}\mbox{
K}$, $\alpha_0=10^{-6}$ and $\Omega_0=\Omega_K$, where
$\Omega_K=(2/3)\sqrt{\pi G\bar{\rho}}=5612 \mbox{ s}^{-1}$ is the
Keplerian angular velocity at which the star starts shedding mass
at the equator (these initial values will be used throughout the
paper). Then, in the first moments of the evolution, bulk
viscosity dominates the dynamics of
Eqs.~(\ref{eq-dif-omega(t)-c-visc}) and
(\ref{eq-dif-alpha(t)-c-visc}). But this lasts just a fraction of
a second. Indeed, the temperature of the star falls so rapidly
that the gravitational-radiation driving effect almost immediately
dominates over the bulk viscosity damping effect. This
preponderance of gravitational radiation continues during most of
the evolution, ending between $t_b=3.6\times 10^6 \mbox{ s}$ (for
$K=-5/4$) and $t_b=7.1\times 10^6 \mbox{ s}$ (for $K\gg 1$) .
Afterwards, shear viscosity dominates the dynamics of the
evolution. At $t=1~\mbox{ yr}$, which corresponds to a temperature
$T=10^9 \mbox{ K}$, we stop evaluating
Eqs.~(\ref{eq-dif-omega(t)-c-visc}) and
(\ref{eq-dif-alpha(t)-c-visc}), since at this temperature
superfluid effects are expected to become important, rendering
invalid our assumptions about the viscous timescale \cite{LM1995}.

\begin{figure}[h]
\includegraphics[width=8.6 cm]{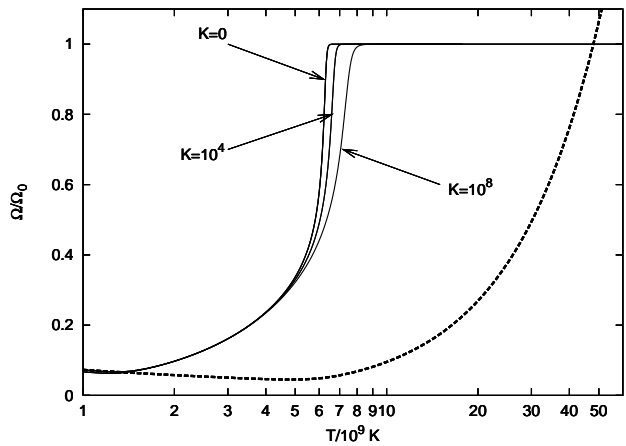}
\caption{\label{fig:omega-temp} Angular velocity of the star as a
function of the temperature. As the star cools and emits
gravitational radiation, its angular velocity decreases to values
which are quite insensitive to the value of $K$. The dotted line
represents the stability curve, $\tau_{GR}^{-1}(\Omega)+
\tau_V^{-1}(\Omega,T) = 0$, i.e., the set of points for which the
damping effect of viscosity balances exactly the driving effect of
gravitational radiation.}
\end{figure}

In Fig.~\ref{fig:omega-temp} the numerical solution of the system
of equations (\ref{eq-dif-omega(t)-c-visc}) and
(\ref{eq-dif-alpha(t)-c-visc}), for different values of the
constant $K$, is represented in a $(\Omega,T)$ diagram.
As we can see there, for a newly born, hot, rapidly-rotating
neutron star, there is an interval of relevant temperatures and
angular velocities of the star for which the \emph{r}-mode
instability is active. For such values of temperature and angular
velocity (corresponding to $0 \lesssim t < t_b$), the influence of
viscosity on the evolutionary equations
(\ref{eq-dif-omega(t)-c-visc}) and (\ref{eq-dif-alpha(t)-c-visc})
is so small, as compared with the influence of gravitational
radiation, that $\Omega(t)$ and $\alpha(t)$ can be determined, in
a very good approximation, by the system of equations:
\begin{eqnarray}
\frac{d\Omega}{dt} &=& \frac83 (K+2)Q \alpha^2
\frac{\Omega}{\tau_{GR}}, \label{eq-dif-omega(t)-s-visc}
\\
\frac{d\alpha}{dt} &=& -\left[ 1 + \frac43 (K+2)Q \alpha^2 \right]
\frac{\alpha}{\tau_{GR}}. \label{eq-dif-alpha(t)-s-visc}
\end{eqnarray}
Therefore, in what follows, Eqs.~(\ref{eq-dif-omega(t)-s-visc})
and (\ref{eq-dif-alpha(t)-s-visc}), which can be solved
analytically, are used to determine $\Omega(t)$ and $\alpha(t)$,
while Eqs.~(\ref{eq-dif-omega(t)-c-visc}) and
(\ref{eq-dif-alpha(t)-c-visc}), which we solve numerically, are
used just to determine $t_b$, i.e., the time at which the damping
effect of viscosity becomes equal to the driving effect of
gravitational radiation.  As mentioned above, the value of $t_b$,
which depends on $K$, lies in the range $3.6\times 10^6 \mbox{
s}\leqslant t_b \leqslant 7.1\times 10^6 \mbox{ s}$.

As shown in Ref.~\cite{ST2005},
Eqs.~(\ref{eq-dif-omega(t)-s-visc}) and
(\ref{eq-dif-alpha(t)-s-visc}) yield the following solution:
\begin{eqnarray}
& & \hspace{-5mm} -\frac{2}{\tilde{\tau}_{GR}}\left(
\frac{\Omega_0}{\sqrt{\pi G\bar{\rho}}}
\right)^6 \left[ 1+\frac43(K+2)Q\alpha_0^2 \right]^6 t \nonumber \\
& & = \ln \frac{\Omega_0}{\Omega} + \ln
\frac{1+\frac43(K+2)Q\alpha_0^2 -
\frac{\Omega}{\Omega_0}}{\frac43(K+2)Q\alpha_0^2} \nonumber \\
& & \quad +\; \sum_{n=1}^5
\frac{\left[1+\frac43(K+2)Q\alpha_0^2\right]^n}{n} \left[ \left(
\frac{\Omega_0}{\Omega} \right)^n-1 \right],
\label{omega(t)-s-visc}
\\
& & \hspace{-5mm} -\frac{1}{\tilde{\tau}_{GR}}\left(
\frac{\Omega_0}{\sqrt{\pi G\bar{\rho}}} \right)^6 \left[
1+\frac43(K+2)Q\alpha_0^2 \right]^6 t \nonumber
\\
& & =\ln \frac{\alpha}{\alpha_0} + \sum_{n=1}^5
\frac{5!}{2n(5-n)!n!}\left[\frac43(K+2)Q\right]^n \nonumber
\\
& & \quad \times \left( \alpha^{2n}- \alpha_0^{2n} \right).
\label{alfa(t)-s-visc}
\end{eqnarray}

An analysis of the above solution makes it clear that the
evolution of the \emph{r}-mode instability proceeds in two stages.

In the first stage of the evolution, the above solution is well
approximated by
\begin{eqnarray}
\frac{\Omega(t)}{\Omega_0} &\approx& 1- \frac43(K+2)Q\alpha_0^2
\exp \left\{ \left( \frac{\Omega_0}{\sqrt{\pi G\bar{\rho}}}
\right)^6
\frac{2t}{|\tilde{\tau}_{GR}|} \right\}, \label{omega-1-fase} \nonumber \\
\\
\alpha(t) &\approx& \alpha_0 \exp \left\{ \left(
\frac{\Omega_0}{\sqrt{\pi G\bar{\rho}}} \right)^6
\frac{t}{|\tilde{\tau}_{GR}|} \right\}, \label{alpha-1-fase}
\end{eqnarray}
i.e., both the angular velocity of the star and the amplitude of
the mode evolve on the gravitational-radiation timescale.

In the second stage of evolution, the solution
(\ref{omega(t)-s-visc}) and (\ref{alfa(t)-s-visc}) is well
approximated by
\begin{eqnarray}
\hspace{-4mm} \frac{\Omega(t)}{\Omega_0} &\approx&  0.63 \left(
\frac{\Omega_0}{\sqrt{\pi G\bar{\rho}}} \right)^{-6/5} \left(
\frac{t}{|\tilde{\tau}_{GR}|} \right)^{-1/5}, \label{omega-2-fase}
\\
\hspace{-4mm} \alpha(t) &\approx& \frac{3.56}{\sqrt{K+2}} \left(
\frac{\Omega_0}{\sqrt{\pi G\bar{\rho}}} \right)^{3/5} \left(
\frac{t}{|\tilde{\tau}_{GR}|} \right)^{1/10}, \label{alpha-2-fase}
\end{eqnarray}
i.e., due to nonlinear effects (the presence of differential
rotation), the amplitude of the mode saturates at a value that
depends crucially on the parameter $K$, namely,
$\alpha_{sat}\propto (K+2)^{-1/2}$, and the angular velocity
decreases to a final value which is quite insensitive to the value
of $K$ (it depends on $K$ through $t_b$).

The smooth transition between the two stages of evolution occurs a
few hundred seconds after the instability becomes active. The
moment of occurrence of this transition can be defined more
precisely by introducing the transition time $t_a$, corresponding
to the moment, determined by the condition
$d^2\alpha/dt^2(t_a)=0$, when the mode's amplitude changes from an
exponential to a much slower power-law growth. Using
Eqs.~(\ref{gr-timescale}), (\ref{eq-dif-omega(t)-s-visc}) and
(\ref{eq-dif-alpha(t)-s-visc}), the above condition yields
$\alpha(t_a)=[12(K+2)Q]^{-1/2}$ or, using the relation $\Omega
\approx \Omega_0 [1+\frac43(K+2)Q\alpha^2]^{-1}$,
$\Omega(t_a)=0.9\Omega_0$. Inserting $\alpha(t_a)$ into
Eq.~(\ref{alfa(t)-s-visc}), or $\Omega(t_a)$ into
Eq.~(\ref{omega(t)-s-visc}), and taking into account that
$4(K+2)Q\alpha_0^2/3 \ll 1$ and $\alpha_0\ll\alpha(t_a)$, one
obtains
\begin{eqnarray}
t_a\approx [521.0-18.5\ln(K+2)]\mbox{ s}.
\end{eqnarray}

\section{\label{sect-detectability} Detectability of gravitational
waves from the \emph{r}-mode instability}

The gravitational radiation emitted by a newly born neutron star,
as it spins down due to the \emph{r}-mode instability, could in
principle be detected by the laser interferometer detectors LIGO
and Virgo using appropriate detecting strategies.

The gravitational wave amplitude, averaged over source and
detector orientation, is given by \cite{OLCSVA1998}
\begin{eqnarray}
|h(t)|=1.3\times 10^{-24} \alpha(t) \left(
\frac{\Omega(t)}{\Omega_K} \right)^3 \left( \frac{20\mbox{
Mpc}}{D} \right), \label{h(t)}
\end{eqnarray}
where $D$ is the distance to the source.

\begin{figure}[h]
\includegraphics[width=8.6 cm]{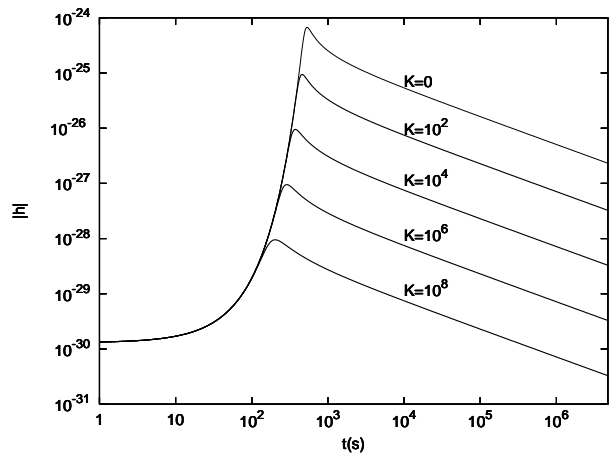}
\caption{\label{fig:h-time} Gravitational wave amplitude $|h(t)|$
for different values of $K$ and distance to the source $D=20\mbox{
Mpc}$. At $t_*\approx[543.1-18.5\ln(K+2)]\mbox{ s}$ the
gravitational wave amplitude reaches its maximum value,
$h_{max}=9.5\times 10^{-25}(K+2)^{-1/2}$.}
\end{figure}

In Fig.~\ref{fig:h-time} the gravitational wave amplitude is shown
for different values of $K$ and a distance to the source of
$D=20\mbox{ Mpc}$. During the first stage of evolution, $|h(t)|$
grows exponentially [see Eqs.~(\ref{omega-1-fase}) and
(\ref{alpha-1-fase})], while during the second stage of evolution
it decreases slowly, as $t^{-1/2}$ [see Eqs.~(\ref{omega-2-fase})
and (\ref{alpha-2-fase})]. The maximum value
$h_{max}\equiv|h(t_*)|$ is achieved when $d|h|/dt(t_*)=0$, which,
using Eqs.~(\ref{eq-dif-omega(t)-s-visc}) and
(\ref{eq-dif-alpha(t)-s-visc}), gives the mode's amplitude
$\alpha(t_*)=\sqrt{3/[20(K+2)Q]}$, or, using the relation $\Omega
\approx \Omega_0 [1+\frac43(K+2)Q\alpha^2]^{-1}$, the star's
angular velocity $\Omega(t_*)=(5/6)\Omega_K$. Substituting these
values of $\alpha$ and $\Omega$ into Eq.~(\ref{h(t)}), we obtain
\begin{eqnarray}
h_{max}=\frac{9.5\times 10^{-25}}{\sqrt{K+2}}  \left(
\frac{20\mbox{ Mpc}}{D} \right). \label{h-max}
\end{eqnarray}
If $\alpha(t_*)$ is inserted into Eq.~(\ref{alfa(t)-s-visc}), or
$\Omega(t_*)$ into Eq.~(\ref{omega(t)-s-visc}),we obtain that the
gravitational wave amplitude reaches its maximum value at
$t_*\approx[543.1-18.5\ln(K+2)]\mbox{ s}$, i.e., shortly after the
transition between the first and second stages of evolution of the
\emph{r}-mode instability.

Let us now derive an analytical expression for the
frequency-domain gravitational wave amplitude $|\tilde{h}(f)|$.
Using the stationary phase approximation, namely,
$|\tilde{h}(f)|=|h(t)|/\sqrt{|df/dt|}$, where $df/dt$ is
straightforwardly obtained from
Eq.~(\ref{eq-dif-omega(t)-s-visc}),
\begin{eqnarray}
\frac{df}{dt}&=& -8.1(K+2)\alpha^2(t) \left( \frac{f}{f_{max}}
\right)^7 \mbox{Hz}/\mbox{s}, \label{dfdt}
\end{eqnarray}
we can easily obtain the frequency-domain gravitational wave
amplitude $|\tilde{h}(f)|$,
\begin{eqnarray}
|\tilde{h}(f)| = \frac{4.6\times 10^{-25}}{\sqrt{K+2}}
\sqrt{\frac{f_{max}}{f}} \left( \frac{20 \mbox{ Mpc}}{D} \right)
\mbox{ Hz}^{-1},
 \label{htilde}
\end{eqnarray}
where $f=2\Omega/(3\pi)$ is the frequency of the emitted
gravitational wave.

Equation~(\ref{htilde}) applies to both stages of evolution, for
frequencies ranging from $f_{min}=(77-80) \mbox{ Hz}$ to
$f_{max}=1191\mbox{ Hz}$, where the maximum frequency corresponds
to the initial value of the angular velocity of the star,
$\Omega_K=5612\mbox{ s}^{-1}$, and the minimum frequency
corresponds to the final angular velocity of the star,
$\Omega(t_b)$, which lies between $0.065 \Omega_K$ (for $K=-5/4$)
and $0.067 \Omega_K$ (for $K\gg 1$).

Let us mention that in Ref.~\cite{OLCSVA1998} an expression for
the frequency-domain gravitational wave amplitude $|\tilde{h}(f)|$
was derived based on the assumption that $dJ/df\propto I$, where
$J$ is the total angular momentum of the star, $f$ is the
frequency of the emitted gravitational wave and $I$ is the moment
of inertia of the unperturbed star. Since within the model of
Ref.~\cite{OLCSVA1998}, the condition $dJ/df\propto I$ only
applies during the second stage of evolution, the expression
obtained there for $|\tilde{h}(f)|$ is valid just for this stage
of evolution. Within our model, however, it can be easily shown
that the condition $dJ/df\propto I$ holds, not only during the
second stage of evolution, but also during the first stage.
Indeed, from Eqs.~(\ref{phys-ang-mom}) and (\ref{total-ang-mom}),
we obtain that
\begin{eqnarray}
\frac{dJ}{d\Omega}=I \left[ 1+ \frac13 \left( 4K+5 \right) Q
\alpha^2 \left( 1+2\frac{\Omega}{\alpha} \frac{d\alpha}{d\Omega}
\right) \right]. \label{dJdf}
\end{eqnarray}
Now we can use Eqs.~(\ref{eq-dif-omega(t)-s-visc}) and
(\ref{eq-dif-alpha(t)-s-visc}) to relate $d\Omega$ and $d\alpha$,
namely,
\begin{eqnarray}
\frac{d\Omega}{\Omega} =
-\frac{\frac83(K+2)Q\alpha}{1+\frac43(K+2)Q\alpha^2}d\alpha.
\end{eqnarray}
Inserting this expression into Eq.~(\ref{dJdf}) and using
$f=2\Omega/(3\pi)$ we obtain
\begin{eqnarray}
\frac{dJ}{df}=\frac{9\pi I}{8(K+2)}. \label{conditionOwenetal}
\end{eqnarray}
Since the above condition holds during both stages of the
evolution, we could use it to derive, in the manner proposed in
Ref.~\cite{OLCSVA1998}, the frequency-domain gravitational wave
amplitude given by Eq.~(\ref{htilde}).

\begin{figure}[h]
\includegraphics[width=8.6 cm]{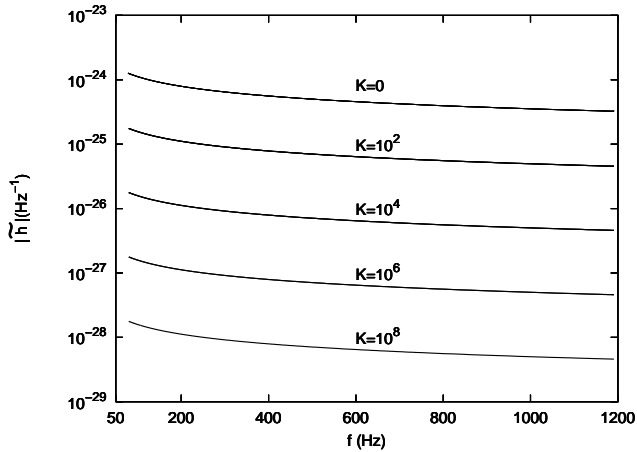}
\caption{\label{fig:htilde-freq} Gravitational wave amplitude in
the frequency domain, $|\tilde{h}(f)|$, for different values of
$K$ and distance to the source $D=20\mbox{ Mpc}$.}
\end{figure}

The gravitational wave amplitude $|\tilde{h}(f)|$ is represented
in Fig.~\ref{fig:htilde-freq} for different values of $K$ and for
gravitational-wave frequencies lying between $f_{min}$ and
$f_{max}$.

It is worth mentioning that within the model of evolution proposed
in Ref.~\cite{OLCSVA1998}, the frequency-domain gravitational wave
amplitude has a spike at high frequencies, due to the fact that
during the first stage of evolution the angular velocity of the
star evolves very slowly on the viscous timescale, leading to a
quasi-monochromatic gravitational wave emission during the first
500 s of evolution. However, as we have seen, if one takes into
account the influence of differential rotation, namely, the fact
that it contributes to the physical angular momentum of the
\emph{r}-mode perturbation, then the angular velocity of the star
evolves in the gravitational-radiation timescale already in the
first stage of evolution [see Eq.~(\ref{eq-dif-omega(t)-c-visc})].
As a consequence, the gravitational wave amplitude
$|\tilde{h}(f)|$ in the first stage of evolution is also given by
Eq.~(\ref{htilde}) and we observe no spike.

Let us now analyze the possibility of detecting the emitted
gravitational waves with laser interferometer detectors LIGO and
Virgo. Because of the complexity of real neutron stars, our
knowledge of the evolution of the \emph{r}-mode instability is
insufficient to predict the gravitational waveform with such an
accuracy that a matched filtering signal-processing technique
would be feasible. However, we can use matched filtering in order
to estimate the detectability of the gravitational-wave signal.
For that purpose, the characteristic amplitude of the signal,
\begin{eqnarray}
 h_c(f)&\equiv&f |\tilde{h}(f)| \nonumber \\
 &=&  \frac{5.5\times 10^{-22}}{\sqrt{K+2}}
\sqrt{\frac{f}{f_{max}}} \left( \frac{20 \mbox{ Mpc}}{D} \right),
\end{eqnarray}
is compared with the rms strain noise in the detector,
\begin{eqnarray}
 h_{rms}(f)\equiv\sqrt{f S_h(f)},
\end{eqnarray}
where $S_h(f)$ is the noise power spectral density of the
detector.

For frequencies in the interval $50\mbox{ Hz}\leqslant f\leqslant
1200\mbox{ Hz}$, the curves for the noise power spectral densities
of the initial LIGO \cite{LIGO}, Virgo \cite{Virgo} and advanced
LIGO \cite{AdvLIGO} detectors are well approximated by the
following analytical expressions, respectively:
\begin{eqnarray}
S_h(f)=S_1\left[ \left( \frac{f_1}{f} \right)^4 +\left(
\frac{f}{f_1} \right)^2 \right], \label{shfligo}
\end{eqnarray}
where $S_1=3.4\times10^{-46}\mbox{ Hz}^{-1}$ and $f_1=142.0\mbox{
Hz}$,
\begin{eqnarray}
S_h(f)=S_2\left[ 1+ \frac16 \left( \frac{f_2}{f} \right)^2 +
\frac16 \left( \frac{f}{f_2} \right)^2 \right], \label{shfvirgo}
\end{eqnarray}
where $S_2=1.5 \times10^{-45}\mbox{ Hz}^{-1}$ and $f_2=249.6\mbox{
Hz}$, and
\begin{eqnarray}
S_h(f) &=& S_3\left\{ 1+ \left( \frac{f_3}{f} \right)^7 -
\frac{10}{3} \left( \frac{f}{f_4} \right) \left[ 1- \left(
\frac{f}{f_4} \right) \right. \right.  \nonumber \\
& & \qquad + \left. \left. \frac{3}{50} \left( \frac{f}{f_4}
\right)^2 \right] \right\}, \label{shfadvligo}
\end{eqnarray}
where $S_3=2.2 \times10^{-47}\mbox{ Hz}^{-1}$, $f_3=52.8\mbox{
Hz}$ and $f_4=421.3\mbox{ Hz}$.

\begin{figure}[h]
\includegraphics[width=8.6 cm]{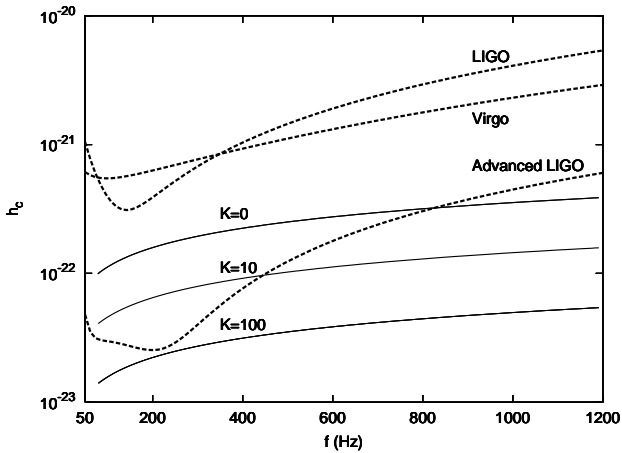}
\caption{\label{fig:hcharac-hrms} The dotted lines correspond to
the rms strain noises in the initial LIGO, VIRGO and advanced LIGO
detectors. The solid lines correspond to the characteristic
amplitude for different values of $K$. The distance to the source
is taken here to be $D=20\mbox{ Mpc}$.}
\end{figure}

In Fig.~\ref{fig:hcharac-hrms} the curves corresponding to the rms
strain noises in the initial LIGO, Virgo and advanced LIGO
detectors are compared with the curves corresponding to the
characteristic amplitude of the gravitational wave signal for
different values of $K$ and for a distance to the source of $D=20
\mbox{ Mpc}$. For such a distance, which includes the Virgo
cluster of galaxies, a few supernovae per year are expected.

The most striking feature of Fig.~\ref{fig:hcharac-hrms} is that
the detectability of gravitational waves from the \emph{r}-mode
instability of newly born neutron stars is drastically reduced as
the initial value of differential rotation associated with
\emph{r}-modes increases. Indeed, for $K\gtrsim100$, even the
advanced LIGO detector would not have enough sensitivity to see
such sources (for $D=20\mbox{ Mpc}$).

The visual comparison, in Fig.~\ref{fig:hcharac-hrms}, between the
characteristic amplitude of the signal and the rms strain noise in
the detector gives us a qualitative measure of the signal-to-noise
ratio for matched filtering. A quantitative determination of the
signal-to-noise ratio is obtained from
\begin{eqnarray}
\left( \frac{S}{N} \right)^2=2\int\limits_{f_{min}}^{f_{max}}
\frac{df}{f} \left( \frac{h_c}{h_{rms}} \right)^2, \label{snr}
\end{eqnarray}
which, for $f_{min}=(77-80)\mbox{ Hz}$ and $f_{max}= 1191\mbox{
Hz}$, yields
\begin{eqnarray}
\frac{S}{N} \approx\frac{1}{\sqrt{K+2}} \frac{20 \mbox{
Mpc}}{D}\times
   \left\{ \begin{array}{rl}
      0.9 & (\mbox{initial LIGO}) \\
      0.7 & (\mbox{Virgo}) \\
     12.9 & (\mbox{advanced LIGO})
    \end{array} \right. \! \! \! \! .
    \label{snr2}
\end{eqnarray}

For the initial LIGO and Virgo detectors, even for small initial
differential rotation ($K\approx0$), the signal-to-noise ratio is
not significant for $D=20 \mbox{ Mpc}$. Of course, this ratio can
be increased if we consider sources located at smaller distances,
but at the cost of decreasing the number of expected supernova
events per year and, hence, the probability of a detection.

\begin{figure}[h]
\includegraphics[width=8.6 cm]{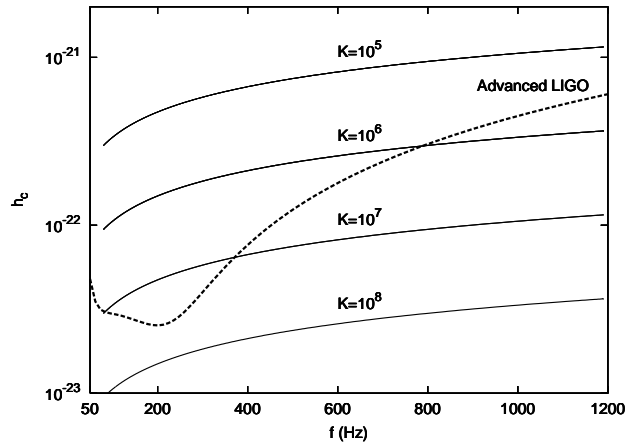}
\caption{\label{fig:hcharac-hrms-30kpc} The dotted line
corresponds to the rms strain noise in the advanced LIGO detector.
The solid lines correspond to the characteristic amplitude for
different values of $K$. The distance to the source is taken here
to be $D=30\mbox{ kpc}$.}
\end{figure}

For the advanced LIGO detector the situation improves: for small
values of $K$, the signal-to-noise ratio is significant even for
$D=20 \mbox{ Mpc}$. Therefore, and since within such a distance
several supernovae per year are expected, one could hope that
advanced LIGO detectors would see gravitational radiation from the
\emph{r}-mode instability of young neutron stars. However, such
hope is based on the assumption that neutron stars are born with
small initial differential rotation associated with
\emph{r}-modes. But this may not be the case. And if a neutron
star is born with substantial differential rotation associated
with \emph{r}-modes (say $K\gtrsim100$), then the emitted
gravitational waves will not be seen even by advanced LIGO. Of
course, as already mentioned above, one could consider smaller
distances to the source, but then the number of events per year
would decrease, decreasing the probability of a detection. In
Fig.~\ref{fig:hcharac-hrms-30kpc} the characteristic amplitude for
high values of $K$ is compared with the rms strain noise in the
advanced LIGO detector for a distance to the source of $30 \mbox{
kpc} $, i.e., within our Galaxy, in which about 2 supernovae are
expected every hundred years \cite{D2006}.

\begin{figure}[h]
\includegraphics[width=8.6 cm]{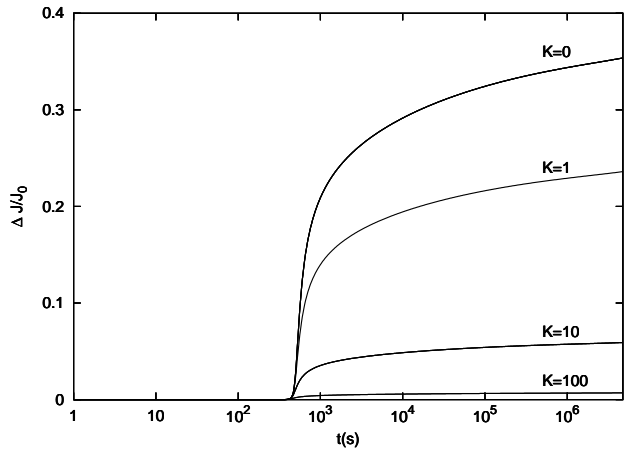}
\caption{\label{fig:ang-mom-away} Angular momentum carried away by
gravitational waves, $\Delta J$, as a function of time for
different values of $K$. For $K\gg1$, only a small part of the
initial angular momentum of the star is carried away by
gravitational waves.}
\end{figure}

The fact that an increase of the initial amount of differential
rotation associated with \emph{r}-modes makes it more difficult to
detect gravitational waves from these modes (for a given distance
to the source ) can be easily understood from angular momentum
considerations. Indeed, it was shown in Ref.~\cite{OL2002}, based
on an unpublished general argument of Blandford, that the
signal-to-noise ratio given by Eq.~(\ref{snr}) is well
approximated by
\begin{eqnarray}
    \left( \frac{S}{N} \right)^2\approx \frac{2G}{5\pi c^3 D^2}
    \frac{\Delta J}{h_{rms}^2},
    \label{snr-delta-J}
\end{eqnarray}
where $\Delta J\equiv J_0-J(t_b)$ is the total amount of angular
momentum carried away by gravitational waves, $J(t)$ is the total
angular momentum of the star  and $J_0\approx I\Omega_0$ is the
initial angular momentum of the star. Using
Eqs.~(\ref{phys-ang-mom}) and (\ref{total-ang-mom}), the total
amount of angular momentum carried away by gravitational waves can
be written as
\begin{eqnarray}
 \frac{\Delta J}{J_0}\approx
 1-\frac{\Omega(t_b)}{\Omega_0}-\frac13 (4K+5)Q
 \frac{\Omega(t_b)}{\Omega_0} \alpha^2(t_b),
\end{eqnarray}
or, since $\Omega\approx\Omega_0[1+\frac43(K+2)Q\alpha^2]^{-1}$,
as
\begin{eqnarray}
\frac{\Delta J}{J_0}  \approx  \frac{3}{4(K+2)}\left(
1-\frac{\Omega(t_b)}{\Omega_0} \right),
\end{eqnarray}
where $\Omega(t_b)=(0.065-0.067)\Omega_0$. Thus, for $K\gg1$, only
a small part of the initial angular momentum of the star is
carried away by gravitational waves\footnote{As shown in
Ref.~\cite{ST2005}, for $K\gg1$, most of the angular momentum of
the unperturbed star is transferred to the \emph{r}-mode
perturbation, due to the fact that the fluid develops a strong
differential rotation. Indeed, after a few hundred seconds of
evolution of the \emph{r}-mode instability, the average
differential rotation increases rapidly, saturating at high values
relatively to the initial angular velocity of the star.} (see
Fig.~\ref{fig:ang-mom-away}) and, consequently, detection of these
waves becomes a more difficult task.

\section{\label{sect-conclusions}Conclusions}

In this paper we have investigated the influence of differential
rotation on the detectability of gravitational radiation emitted
by a newly born, hot, rapidly-rotating neutron star, as it spins
down due to the \emph{r}-mode instability.

A model of evolution of the \emph{r}-mode instability that takes
into account differential rotation \cite{ST2005} has been used to
derive the gravitational wave amplitude $|h(t)|$ and its Fourier
transform $|\tilde{h}(f)|$. We have shown that the maximum value
of the gravitational wave amplitude, $h_{max}$, depends on the
amount of differential rotation at the time the \emph{r}-mode
instability becomes active, namely, $h_{max}\propto(K+2)^{-1/2}$.
We have also shown that the frequency-domain gravitational wave
amplitude, $|\tilde{h}(f)|$, has no spike at high frequencies, as
opposed to the results of Ref.~\cite{OLCSVA1998}. There, the spike
of $|\tilde{h}(f)|$ is due to the fact that during the first stage
of evolution of the \emph{r}-mode instability the angular velocity
of the star evolves very slowly on the viscous timescale, leading
to a quasi-monochromatic gravitational wave emission during the
first few hundred seconds of evolution. However, as we have seen,
if one takes into account the influence of differential rotation,
then the angular velocity of the star evolves in the
gravitational-radiation timescale already in the first stage of
evolution and, as a consequence, during this stage of evolution
the gravitational wave is not monochromatic and $|\tilde{h}(f)|$
has no spike.

We have assumed matched filtering in order to investigate the
detectability of the gravitational wave signal. However, our
knowledge of the evolution of the \emph{r}-mode instability is
insufficient to predict the gravitational waveform with such an
accuracy that an optimal matched filtering signal-processing
technique would be feasible. But non-optimal signal-processing
strategies could be developed such that they would yield results
not much different from the ones obtained with matched filtering
(see, for instance, the hierarchical search strategies proposed in
\cite{BC2000}). Therefore, our results concerning the
detectability of gravitational waves from \emph{r}-modes can be
considered to be a good approximation.

Assuming matched filtering, the characteristic amplitude of the
signal $h_c(f)$ is compared with the rms strain noise in the
initial LIGO and Virgo detectors, as well as with the expected rms
strain noise in the advanced LIGO detector [see
Figs.~\ref{fig:hcharac-hrms} and \ref{fig:hcharac-hrms-30kpc} and
Eq.~(\ref{snr2})]. We conclude that the detectability of
gravitational waves from the \emph{r}-mode instability of newly
born neutron stars is drastically reduced as the initial value of
differential rotation associated with \emph{r}-modes increases.
For the initial LIGO and Virgo detectors, the signal-to-noise
ratio obtained with matched filtering for sources located at 20
Mpc is smaller than unity even for $K\approx 0$. For advanced
LIGO, if neutron stars are born with significant differential
rotation associated with \emph{r}-modes, then detection of
gravitational waves would be possible only if the distance to
these neutron stars is considerably smaller than 20 Mpc. For
instance, if $K=10^5$, a signal-to-noise ratio greater than 10
requires a source located no more than about 80 kpc away [see
Eq.~(\ref{snr2})], i.e., within a sphere containing the Milky Way,
the Magellanic Clouds and a few more small galaxies, in which just
a few supernovae per century are expected. However, if the initial
value of differential rotation associated with \emph{r}-modes is
small ($K\approx0$), then $S/N\gtrsim 10$ can be obtained for
$D=20 \mbox{ Mpc}$. Since within a volume of radius 20 Mpc several
supernovae per year are expected, it is quite possible that
advanced LIGO will detect gravitational radiation from the
\emph{r}-mode instability of young neutron stars.

\begin{acknowledgments}
This work was supported in part by the Funda\c c\~ao para a
Ci\^encia e a Tecnologia (FCT), Portugal.
\end{acknowledgments}

\end{document}